\documentclass[usenatbib]{mn2e}
  
\newcommand{\lta}{\mbox{\small\raisebox{-0.6ex}{$\,\stackrel
{\raisebox{-.2ex}{$\textstyle <$}}{\sim}\,$}}}
   
\title{First Results of the 74 MHz VLA-Pie Town Link. \\  Hercules A at Low
Frequencies}

\author[Nectaria A. B.  Gizani, A. Cohen and N. E. Kassim]{Nectaria
A. B. Gizani,$^{1,2}$~\footnote{Current address: Departamento de 
F$\acute{i}$sica, 
Faculdade Ci\^encias, Universidade de Lisboa, Edif$\acute{i}$cio C8, 
Campo Grande, 
1749-016 Lisboa, Portugal},
A. Cohen,$^3$ and N. E. Kassim,$^3$ \\ $^1$ Grupo de Astrof\'{i}sica da 
Universidade de Coimbra e Observat\'{o}rio
Astron\'{o}mico da Universidade de Coimbra,\\ Santa Clara, 3040 Coimbra, 
Portugal\\$^2$Institute of Astronomy and
Astrophysics, National Observatory of Athens, I. Metaxa \& B. Pavlou,
Lofos Koufou, Palaia Penteli,\\ $^{~}$ 15236 Athens, Greece.\\ $^3$
Naval Research Laboratory, Code 7213, Washington, DC 20375, USA\\}

\begin{document}

\bibliographystyle{mn2e} 
\maketitle

\begin{abstract}

We present the results of the first successful observations of the Pie
Town link with the Very Large Array (VLA) at 74~MHz on Hercules A.
The improvement in resolution from 25 arcsec to 10 arcsec resolves the
helical- and ring-like features seen at higher frequencies.  We also
present new high dynamic range images of this powerful radio galaxy at
325~MHz. Our low frequency observations confirm the multiple outburst
interpretation of the spectral index differences at high
frequencies. Comparison between our radio and ROSAT X-ray data does
not reveal any association between the X-ray emission from the cluster
and the radio lobes. There are no extra regions of radio emission at
74~MHz.

\end{abstract}

 
\section{Introduction}

\subsection{Hercules A}

Hercules A (Her~A) is a complex extended radio galaxy at a low
redshift of $ z=0.154$ (see \citealt{GL2004,GL2003} for a review on
this source and results from new observations).  It is one of the most
luminous radio sources in terms of apparent and intrinsic brightness.
Its total power output is nearly as great as Cygnus A.  At low
frequency, it is one of the brightest radio sources in the sky at 170
and 800 Jy at 325 (VLA P-band) and 74~MHz (VLA 4-band) respectively
\citep{1993K}.  Its total radio luminosity is $\sim
3.8\times10^{37}$~W in the band 10~MHz to 100~GHz~\footnote{We use
H$_{\circ}$=65 kms$^{-1}$~Mpc$^{-1}$ and q$_{\circ}$=0 throughout for
consistency with the series of papers on this object. We are not very
wrong in doing this, since the recent WMAP (Wilkinson Microwave
Anisortopy Probe, \citealt{Betal2003}) data give H$_{\circ}$=71
kms$^{-1}$~Mpc$^{-1}$ and q$_{\circ}$=-0.595, which do not change our
results significantly.}. Hercules A is classified as an FR1.5
\citep{Dreher.etal1984}.  With a linear
size of 540 kpc and a width of $\simeq$ 250 kpc (angular size = $194
\times70$~arcsec), this radio galaxy posesses an unusual jet-dominated
morphology, almost symmetrically extended lobes and no compact
hotspots.  The ring-like/helical features on both sides of the radio
emission form an almost symmetrical sequence which suggests successive
ejections from the active nucleus \citep{GL2003}.

Her~A has also been studied in the X-rays with the ROSAT PSPC
(Position Sensitive Proportional Counter) and HRI (High Resolution
Imager) \citep{GL2004}. The cluster is luminous in X-rays with a
bolometric luminosity $L_{\rm bol} = 4.8\times 10^{37}$ W. Her~A
itself lies at the center of the cluster, which should also have a
cooling flow at the center, when the powerful jets are absent,
Currently, however, it is critically disturbed by the expansion of the
radio lobes. The PSPC spectrum reveals a cool component of the
intraluster medium (ICM) with $kT = 2.52$ keV. The total mass of the
cluster is estimated to be 1.5$\times 10^{14}$ M$_{\odot}$.

\citet{GL2003}, have made an extensive study of Her~A with the VLA
1.4~GHz band (L-band, A-, B-, C- configuration) at 1295, 1364.9,
1435.1, and 1664.9~MHz; 8.4~GHz band (X-band, B-, C-, D-
configuration) at 8414.9 and 8464.9~MHz and reprocessed the 5~GHz band
data (C-band, B-, C- and D- configuration) at 4872.6~MHz by
\citet{Dreher.etal1984}. Gizani \& Leahy 2003, have produced a
spectral index map with 1.4 arcsec resolution which provided a
detailed picture of the plasma throughout the complex morphology, and
gave a fairly good idea about the temporal history of the
plasma. Clearly the spectral index map separates the source in
distinct regions of brighter, flatter (younger material) and fainter,
steeper ( older material), which is some evidence for spectral
curvature, suggesting that we may be witnessing a renewed outburst
from the active nucleus \citep{GL2003}. Generally spatial and spectral
variations in a source imply temporal variations as well \citep{Leahy1991}.

However the multifrequency measurements of \citet{GL2003} contained
data from only a relatively small frequency range (1.4 to 8.4~GHz),
with each one of the four frequencies of the 1.4~GHz band, not very
far apart from eachother.  More quantitative conclusions, such as
estimating the aging of Her~A require a much wider frequency range
which extends to low enough frequencies to map the break frequency
throughout this radio galaxy. Low frequencies are more likely to trace
the shape of the injected spectrum, before the aging effects have
introduced spectral breaks. With this goal in mind, we have observed
Her~A at both 325 and 74~MHz.

\begin{figure*}
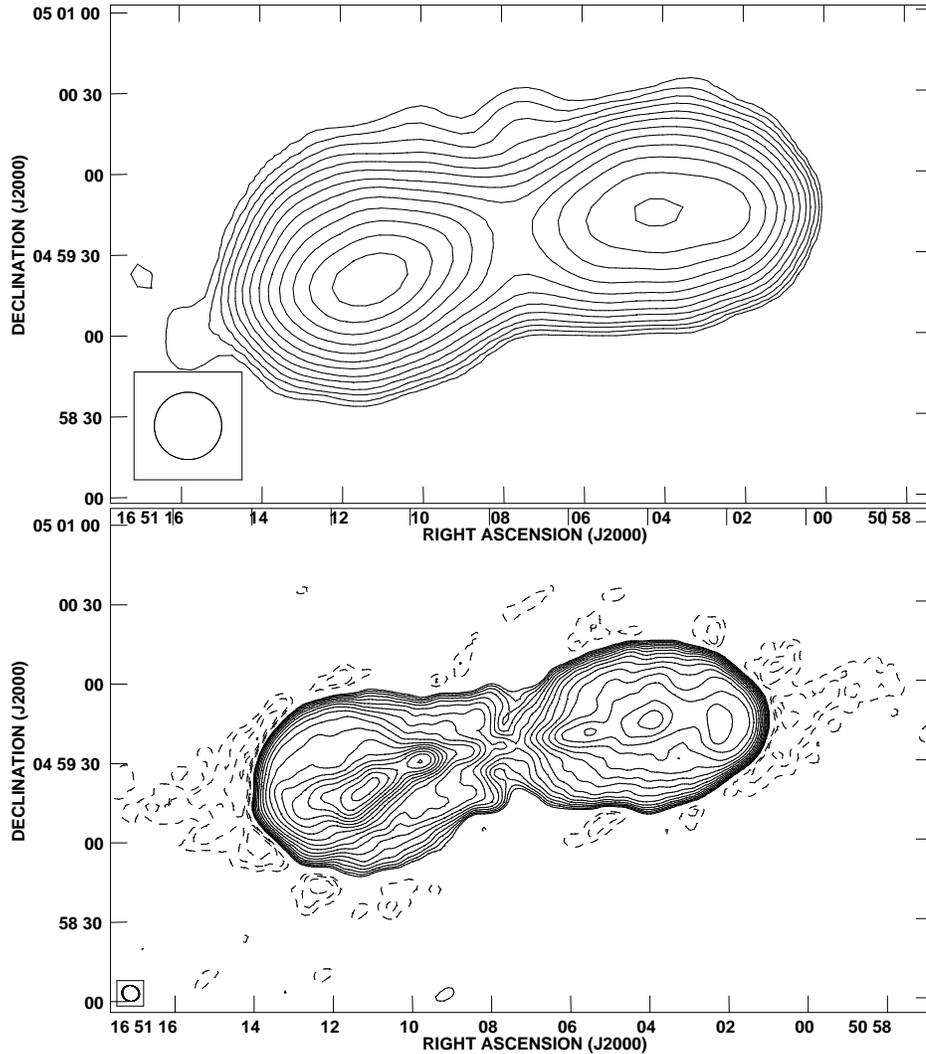

\centering \setlength{\unitlength}{1cm}

\begin{picture}(8,7)
\put(-3,9){\includegraphics{./ngizani1.ps}}
\end{picture}
 
\begin{picture}(8,6.5)
\put(-3,8.8){\includegraphics{./ngizani2.ps}}
\end{picture}

\caption{Top: Snapshot image at 74~MHz A-array with only 8
antennas. The peak brightness is 119 Jy beam$^{-1}$, the rms noise is
0.5 Jy beam$^{-1}$ and the resolution is 25 $\times$ 25
arcsec. Contours are logarithmic separated by factors $\sqrt{2}$. The
first contour is at 1.5 Jy beam$^{-1}$. Bottom: Snapshot image at
325~MHz. The peak brightness is 6.3 Jy beam$^{-1}$, the rms noise is
3.7 mJy beam$^{-1}$ and the beam is 6.47 $\times$ 5.87 arcsec with
position angle PA = 67.5 deg . As before, contours are logarithmic
separated by factors $\sqrt{2}$ starting at 15 mJy beam$^{-1}$. In
both figures coordinates are given in the Julian J(2000) coordinate
system to match our new maps.}
\label{namir} 
\end{figure*}

The archived VLA data, at these frequencies are of low resolution and
especially the 74~MHz data are poor. Fig.~\ref{namir} shows contours
of the old maps of Her~A at 74 (top) and 325~MHz
(bottom). \citet{GL2003} have also found the faint extended emission
'cocoon', at 1.4~GHz, 1.4 arcsec resolution, surrounding the core and
jets of Her~A shown in Fig.~\ref{namir} (see also \citealt{1993K}).

It is clear that even in the A-configuration, the VLA does not have the 
resolution required at 74~MHz to provide a useful spectral index map.  
Therefore, we turn to the most recent advance in high-resolution, 
low-frequency observations, the VLA-Pie Town link, which has just been 
upgraded to operate at 74~MHz.

\subsection{The VLA-Pie Town Link at 74~MHz}

In spite of the importance that low frequencies ($\nu \leq $100~MHz) have
played in the discovery and development of radio astronomy (Jansky
1933, Reber 1940, Ryle \& Smith 1948, Hewish 1952), the region of the
electromagnetic spectrum below 100~MHz remains among the most poorly
explored, despite its great scientific potential. While interferometers such
as MERLIN (Multi
Element Radio Linked Interferometer) have brought higher resolution
interferometery to frequencies as low as 151~MHz (resolution 3 arc-second),
sub-arcminute imaging below 100 MHz has been limited because ionospheric
phase fluctuations have restricted interferometer baselines to $\sim$ 5~km
(see for example \citet{Eric1982} and \citet{Rees1990}). The resulting poor
angular resolution and consequently confusion limited sensitivity have
left frequencies below 100 MHz in a seeming dark ages relative to centimeter
wavelength radio astronomy that can exploit the long baselines
($\geq$35~km) of instruments such as the VLA, MERLIN, and the VLBA.  An
important
breakthrough was the demonstration that the technique of
self-calibration could mitigate the ionospheric limitations and
thereby achieve sub-arcminute resolution imaging at the longer wavelengths
\citep{1993K}. During the 1990s, the National Radio Astronomy Observatory
(NRAO) and Naval Research Laboratory (NRL) developed 74~MHz system at the
VLA was the first connected element imaging array to successfully break the
``ionospheric barrier'' below 100 MHz by achieving $\sim20 $~arcsec
resolution at 74 MHz. In 2001, NRL and NRAO collaborated again to develop a
74~MHz observing system on the Pie Town VLBA antenna (hearafter PT)
located 50~km from the center of the VLA and linked to it via optical
fiber. In 2002 the first successful fringes were
obtained at both 74 and 330~MHz on the radio galaxy 3C\,123, thereby
marking another important step forward in the development of high angular
resolution, low frequency
imaging.

Today, low frequency systems on the VLA and the Giant Metrewave Radio
Telescope (GMRT) are creating a quiet renaissance in long wavelength
radio astronomy, producing high angular resolution, high sensitivity
images \citep{2004K} and stimulating the development of instruments
with even greater capability (e.g. Low Frequency Array (LOFAR), see
\citet{2002K}).

In this paper we present
the first successful scientific observations of a cosmic radio source,
the powerful FR I/II radio galaxy Her~A, using the new VLA+PT system,
thereby achieving the highest angular resolution image ever obtained
by a connected element interferometer at a frequency below 100~MHz.

\subsection{Outline of Paper}

We present the total intensity map of Her~A at 74~MHz employing the
VLA-PT link, the corresponding map at 325~MHz as well as the spectral
index map between 74~ and 325~MHz. The spectral imaging analysis in
the frequency range (0.074~ to 8.4~GHz) and the resulting spectral
aging study will be presented in a subsequent
paper. Section~\ref{obs} details the observations we made and
Section~\ref{danal} describes the data reduction and analysis.
Results are discussed in Section~\ref{res} and in Section~\ref{concl}
we summarise our conclusions.

\section{Observations}
\label{obs}

We have used the NRAO's VLA to obtain total intensity images of Her~A
at both 325 and 74~MHz, in order to make a spectral aging study.  As
the limiting factor in creating an accurate spectral index map of this
source is the low resolution obtainable at 74~MHz, we employed the new
VLA-PT link to increase the resolution at that frequency.  At 325~MHz,
we observed both in the A and B-configurations in order to maintain
both high resolution and sensitivity to the large scale source
structure. At 74 MHz, we are restricted to 1.5 MHz bandwidth, because
that is the only part of the frequency spectrum in that region that is
protected for astronomy.  At 325 MHz we chose the 6.25 MHz bandwidth,
because this is the maximum that the correlator can handle while still
having channels narrow enough to avoid bandwidth smearing.

We observed in the multichannel continuum mode in order to 
reduce the effects of bandwidth smearing and to more efficiently remove
radio frequency interference (RFI).  In the A-configuration, we observed
both frequencies, 325 and 74~MHz, simultaneously, each in a separate IF. 
In the B-configuration, we only observed in the 325~MHz band, and we 
observed in two IF settings (328.5 and 321.5625~MHz).
The observations contained nearly full tracks, necessary to maximise the 
$uv$-coverage so that we could map the complex structure of Her~A.  
Full details of the observations are given in Table~\ref{tobs}. 

\begin{table*}
\caption{Observations of Her~A.}
\begin{center}

\begin{tabular}{ccccccl} \hline
Config & Frequency & Bandwidth & Channels & IFs & Time & Dates \\ 
 & MHz & MHz & \# & \# & hrs & \\ \hline 
A\,+\,PT  &  74  & 1.56  & 64 & 1 & 8 & April 2002 \\
A & 325 & 6.25 & 16 & 1 & 8 & April 2002 \\ 
B & 325 & 6.25 & 16 & 2 & 4 & August 2002 \\ \hline

\end{tabular}
\end{center}

\label{tobs}
\end{table*}

Cygnus A (Cyg A) was used as a bandpass calibrator at both
frequencies.  As Cyg A is resolved at both frequencies, accurate
source models were required for calibration at each frequency, which
are also available online\footnote{\tt
http://lofar.nrl.navy.mil/pubs/tutorial/node49.html}. For the 74~MHz
model of Cyg A, we have used a high resolution ($\sim 9$ arcsec) image
produced with the PT link only about 2 weeks before our own
observations (Kassim, Carilli, Harris and Perley, private
communication). For the 325~MHz model of this source we have made use
of the model available in the on-line low-frequency data reduction
tutorial (see Section~\ref{danal}).

For flux calibrators, we used 3C\,286 and Cyg A for 325 and 74~MHz
respectively. For Cyg A again we used the model, because of the
resolved source structure, restricting to inner channels. For 3C\,286,
which is unresolved at this frequency and resolution, we did not
require a model for calibration. No phase calibrator was observed,
because with the high amount of flux density in Her~A we determined
that self-calibration would be sufficient for this purpose.
Approximately 90 per cent of the time was spent on the target source,
the remainder on calibrators and slewing.

The observations went almost as planned.  Some of the 74~MHz data were
affected by interference and unstable rack temperatures, and data of
the 325~MHz were occasionally lost for other reasons including antenna
motor faults. The total downtime per antenna was 10.3 minutes. 

\section{Data Reduction \& Analysis}
\label{danal}

The greatest difficulty in the data reduction at low frequencies comes
from the non-coplanar array geometry and the large field of view
(f.o.v.) of the VLA. For the reducing and imaging of the data we have
used routines in the Astronomical Image Processing Software (AIPS). We
have largely followed the methodology described by T. J. W. Lazio,
N. E. Kassim, \& R. A. Perley in ``Low-Frequency Data Reduction at the
VLA: A Tutorial for New Users'', also found on-line\footnote{\tt
http://lofar.nrl.navy.mil/pubs/tutorial/}.

After performing bandpass (using Cyg A) and flux calibration (using
Cyg A at 74~MHz and 3\,C286 at 325~MHz),excision of RFI was performed
by first clipping all data points well above the actual flux in the
image.  Our clip levels were 300 and 5,000 Jy/beam for 325~MHz and
74~MHz respectively.  Once the worst RFI was removed from the channel
data, we averaged the channels as far as possible without introducing
bandwidth smearing (aips task {\sc splat}).
%\footnote{\textsc{aips} task \texttt{SPLAT}}.  
At 325~MHz, we averaged the entire bandwidth to 
a single channel per IF.  At 74~MHz, we took the central 56 channels
out of the original 64 and averaged this down to 7 channels.  After
channel averaging, additional RFI flagging was performed by 
hand (aips task {\sc tvflg}).
%\footnote{\textsc{aips} task \texttt{TVFLG}}.  

\subsection{Mapping}
\label{map}

The full f.o.v.  , defined roughly by the primary beam area, has a
diameter of $2.5^\circ$ and $11.5^\circ$ for 325 and 74~MHz
respectively.  However, because Her~A is one of the brightest objects
in the sky, it completely dominates all other sources in the field,
making it unnecessary to clean other sources in the primary beam area.
Therefore, only a small region surrounding Her~A (size $\sim3$ arcmin)
needed to be imaged, removing the need for the usual ``fly's eye''
approximation~\footnote{A ``3--D'' wide-field imaging technique, which
does an interpolation between a set of small overlapping images of
multiple pointings, called facets, producing a single image centered
on the facet centered on the source.}
%the pointing position of the facet
 to the three-dimensional Fourier inversions necessary to map
the entire primary beam area.

\begin{figure*}

\centering
\setlength{\unitlength}{1cm}

\begin{picture}(17.5,12.0) 
\put(-1,13){\includegraphics{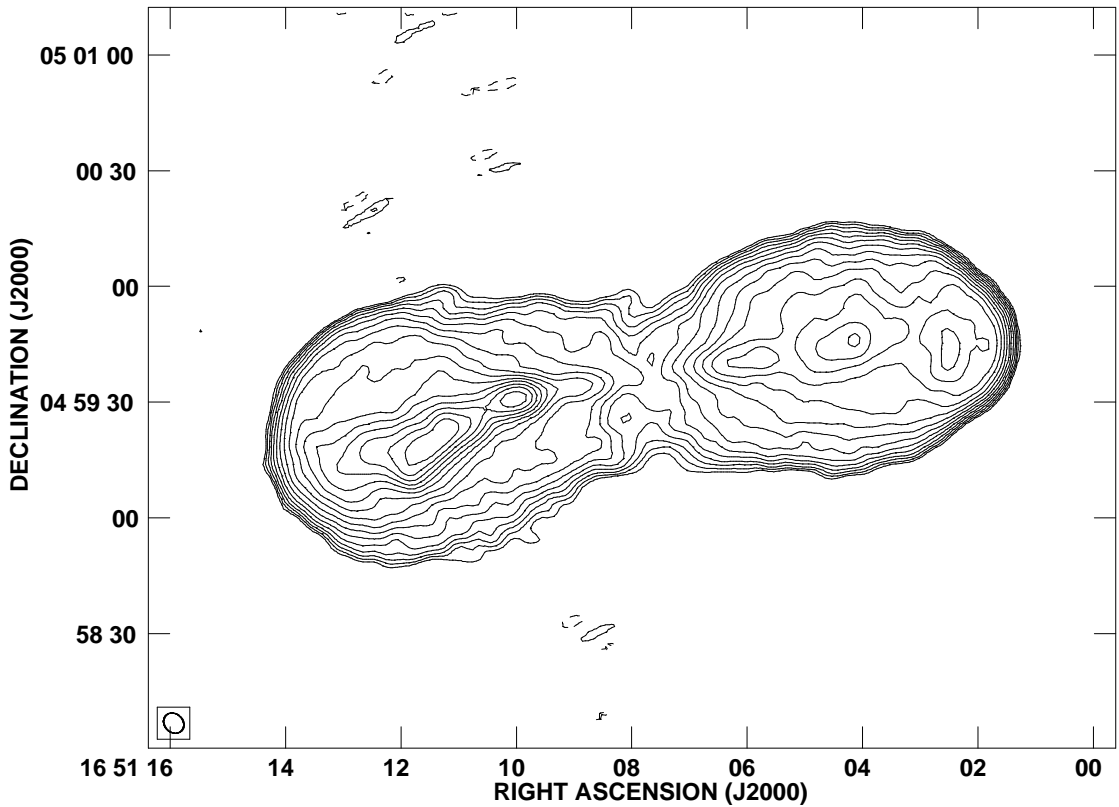}}
\end{picture}

\caption{The 325~MHz map produced by combining both the A and B array
data. The beam size is $5.68 \times4.77 $~arcsec at a position angle
45.17 deg and it is shown in the lower left-hand corner. The dynamic
range in the image is 800:1 and the rms noise is $\simeq$ 8
mJy$bm^{-1}$. Contours are separated by factors $\sqrt{2}$, 
starting at 25 mJy beam$^{-1}$. Coordinates are as in Figure~\ref{namir}. }
\label{Q}
\end{figure*}

\begin{figure*}

\centering
\setlength{\unitlength}{1cm}

\begin{picture}(17.5,12.0) 
\put(-1,13){\includegraphics{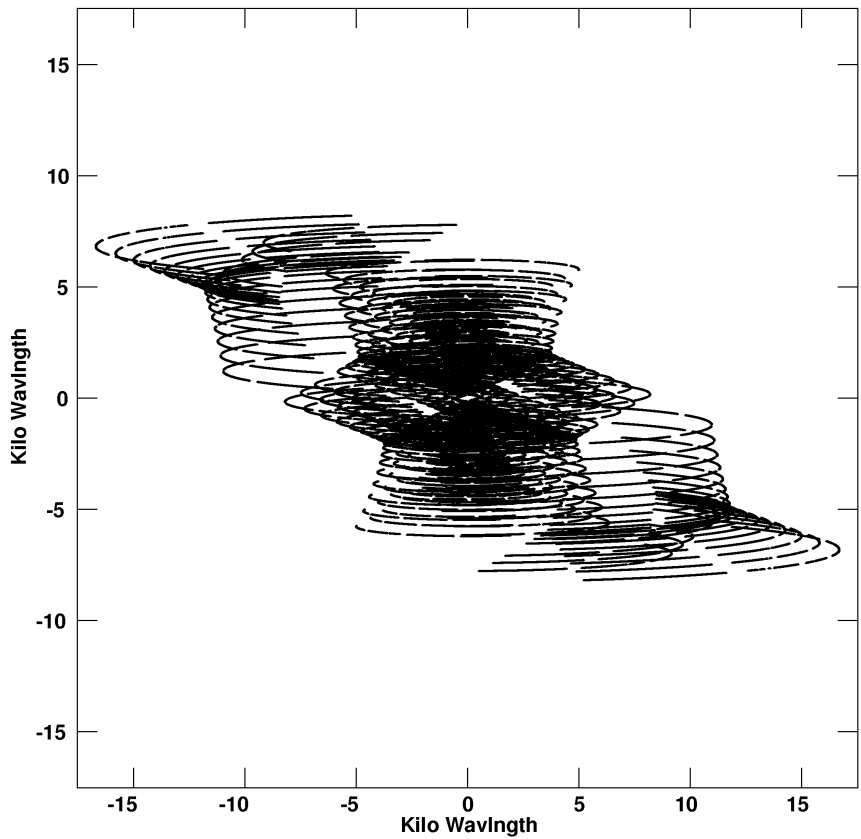}}
\end{picture}

\caption{ The $uv$-coverage for the 74~MHz observations of Her~A utilizing the PT link. 
The
near equatorial location of Her~A causes the long VLA-PT baselines to
be mostly in the east-west direction. As a result, the synthesised
beam for the VLA-PT link image shown in Figure~\ref{74pt} is elongated, while
in the case of all other images we present it is more nearly circular because they 
did not utilize the PT link.
Because the outer regions of the $uv$-plane are
relativley sparsely covered compared to the inner regions, it was
necessary to use nearly uniform weighting to achieve the full
resolving power of the VLA-PT link.  }
\label{uv}
\end{figure*}

\begin{figure*}

\centering
\setlength{\unitlength}{1cm}
  
\begin{picture}(17.5,10.0) 
\put(-1,11){\includegraphics{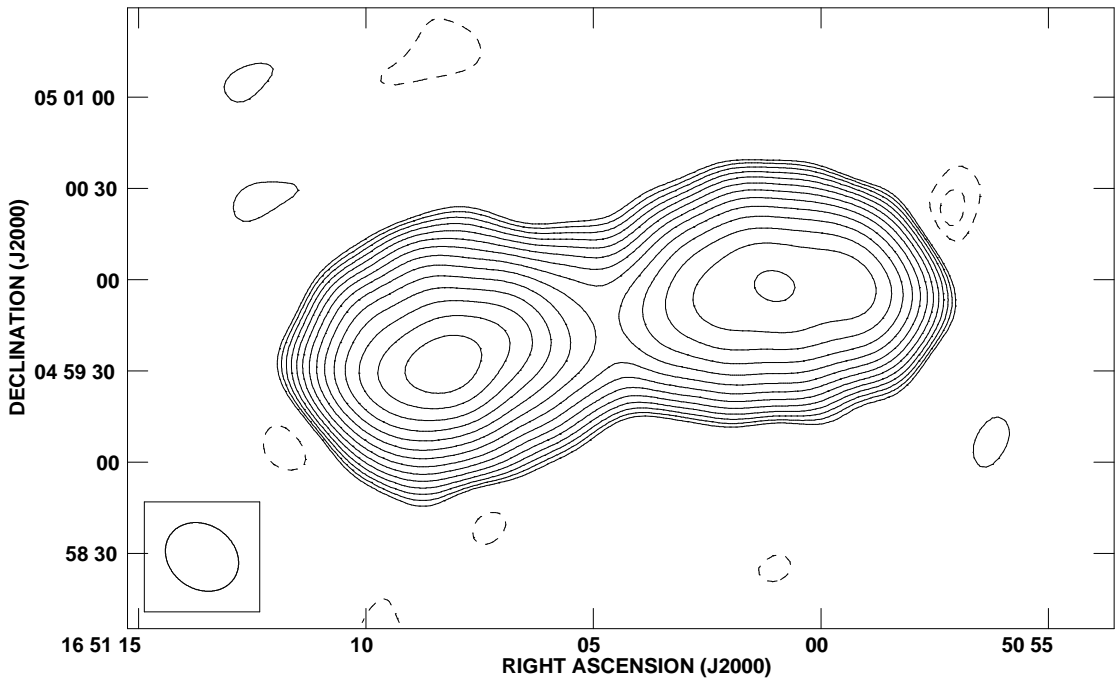}}
\end{picture}

\caption{The 74~MHz map without the PT link. The beam size is (25.4
$\times$ 21.1 arcsec with PA = 55.0 arcsec) shown in the lower
left-hand corner. Contours are separated by factors $\sqrt{2}$, and
starting at 0.5 mJy beam$^{-1}$. Coordinates are as in
Figure~\ref{namir}.}
\label{74}
\end{figure*}

\begin{figure*}

\centering
\setlength{\unitlength}{1cm}
  
\begin{picture}(17.5,12.0) 
\put(-1,14){\includegraphics{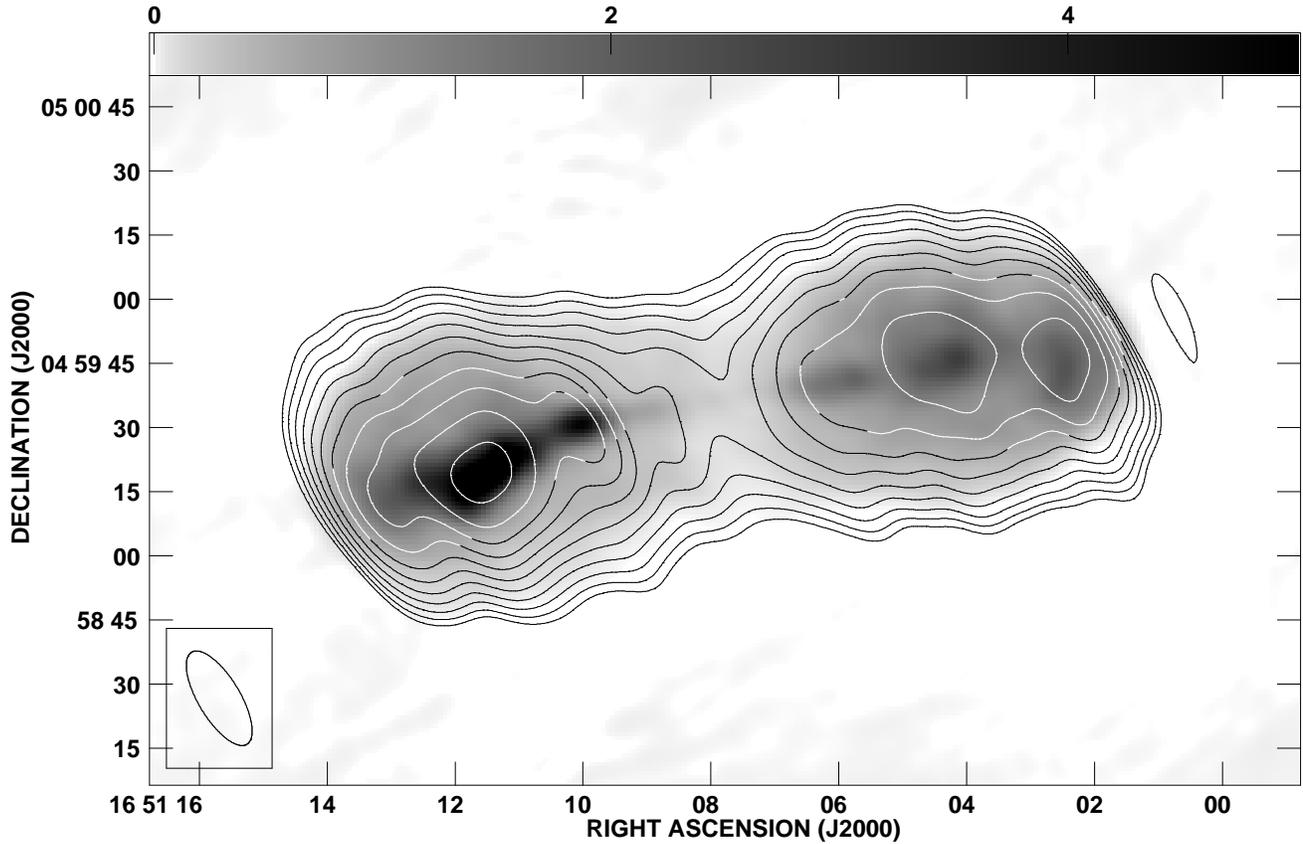}}
\end{picture}

\caption{The best 74~MHz map with the PT antenna (contours) overlayed on top of
the 325~MHz map (grey-scale). The dynamic range in the image is $\sim$ 360:1 with 
rms noise 0.2 Jy$bm^{-1}$. The beam size 
($25.12 \times9.75 $~arcsec, PA = $30.95^\circ$) 
is shown in the lower left-hand corner. Contours are separated
by factors $\sqrt{2}$, and starting at $\pm2$ Jy/beam. Coordinates are 
as in Figure~\ref{namir}. }
\label{74pt}
\end{figure*}

For the 325~MHz data, we combined the A-configuration and B-configuration
data in the $uv$ plane before 
imaging (aips task {\sc dbcon}).
%\footnote{\textsc{aips} task \texttt{DBCON}}.  
For the 74~MHz data, we have included PT link data, which increases 
the maximum baseline to 73~km.  However, the coverage of the outer regions 
of the $uv$-plane, provided by the VLA-PT baselines, is sparse and 
elongated nearly east-west (Fig. \ref{uv}) This is due to the near equatorial 
location of Her~A ($\delta \simeq 5^\circ$), and results in a highly elongated 
synthesised beam.  Imaging was performed for each frequency with a nearly 
uniform weighting (robust factor of $-2$) and with  1.5~arcsec 
pixels (aips task {\sc imagr}). 
%\footnote{\textsc{aips} task \texttt{IMAGR}}.  
Several rounds of phase-only self-calibration were performed on each
data set until the map noise stopped decreasing (aips task {\sc
calib}). No amplitude self-calibration was done.  The final images
produced this way are presented in Figs.~\ref{Q},~\ref{74}
and~\ref{74pt}.
%\footnote{\textsc{aips} task \texttt{CALIB}}.

Fig.~\ref{Q} presents the new 325~MHz image. The angular resolution of
this map is $5.68 \times4.77 $~arcsec at a position angle
$45.17^\circ$. The dynamic range in the image is 830:1 and the rms
noise is $\simeq$ 8 mJy/beam.  Fig.~\ref{74pt} shows the 74~MHz map
with PT overlayed on top of the 325~MHz map. The resolution of the map
is $25.12 \times9.75 $~arcsec at a position angle $30.95^\circ$. The
dynamic range in the image is $\sim$ 385:1 with rms noise 0.2 Jy/beam.
The quality of the 74~MHz map in particular is limited by the fact
that the $uv$ coverage for the PT link is sparse and not very good
near the equator. One can see artifacts near the ends of the lobes in
Fig.~\ref{74pt}, probably due to residual calibration problems.
Fig.~\ref{uv} shows that the PT baselines hardly touch the VLA
ones. Self-calibration cannot get rid of small errors if the baselines
from all telescopes do not cross over each other very densely, and
this may have led to the artifacts in the map. Though the synthesised
beam is highly elongated, the axis of maximum resolution fortunately
is aligned nicely with the axis of the jets in Her~A, allowing many
features to be resolved which were not previously seen at 74~MHz
(c.f. Fig.~\ref{namir}, top).  Fig.~\ref{74} shows the 74~MHz map
without the PT link for comparison.

\section{Discussion} 
\label{res}

\subsection{The Radio Characteristics}

The total radio power of Her~A at 325 and 74~MHz from our observations
is P$_{\rm 325~MHz} = 1.15 \times 10^{27} {\rm \, W Hz^{-1}sr}^{-1}$
and P$_{\rm 74~MHz} = 4.65 \times 10^{27} {\rm \, W Hz^{-1}sr}^{-1}$
respectively using a mean spectral index $\alpha \simeq -1.01$. 
%As
%mentioned in the introduction using the new values for H$_{\circ}$ and
%q$_{\circ}$ suggested by WMAP, our results do not change significantly
%(c.f. P$_{\rm 325~MHz} = 1.07 \times 10^{27} {\rm \, W
%Hz^{-1}sr}^{-1}$ and P$_{\rm 74~MHz} = 4.34 \times 10^{27} {\rm \, W
%Hz^{-1}sr}^{-1}$).

It is clear from comparing Figures \ref{74} and \ref{74pt} that the
addition of the PT-link data has resolved many features in the jets
that had only previously been identified at higher frequencies.  The
extended diffuse emission which forms the `cocoon' is more clearly
defined in the PT-link map.  At 325~MHz, almost all emission was seen
in the old maps within the errors (angular size $199 \times68
$~arcsec).  In the original observations at 74~MHz \citep{1993K} the
source had appeared larger ($227 \times95 $~arcsec) but with the new
PT-link resolution the size is $185 \times71 $~arcsec, which agrees
much better with the 325~MHz result. The flux densities measured from
our new maps are S$_{\rm 325~MHz}$= 206 Jy and S$_{\rm 74~MHz}$=835
Jy.  Direct comparison between Figs.~\ref{74pt}, ~\ref{Q} and
our 1.4~GHz data shows that all radio emission, faint and bright, young
and old, is detected (see also Paper II).

The 325 and 74~MHz images (Figs.~\ref{Q} and ~\ref{74pt}) show many of
the same bright features on both jets as are seen in the higher
resolution 1.4~GHz image.  These are identified
with the helical- and ring-like structures in the eastern and western
side of the radio emission respectively.  These features also have
flatter spectrum than the surrounding medium and so they belong to the
younger material ejected from the nucleus \citep{GL2003}.  Neither map
resolves the core which is found to be very faint and, surprisingly
enough, to have a steep spectrum \citep{GL2003}. Our spectral aging
study will be given in a follow-up paper (Gizani, Cohen, Kassim \&
Leahy, in prep.) where we expect to present a detailed
characterization of the spectral curvature. At the current observing
frequencies we expect the spectrum to be flatter than at higher
frequencies because of the aging of the relativistic particle energy
spectrum.

\subsection{Spectral Index}

\begin{figure*}
\centering
\setlength{\unitlength}{1cm}

\begin{picture}(8,8)
\put(-3,9.5){\includegraphics{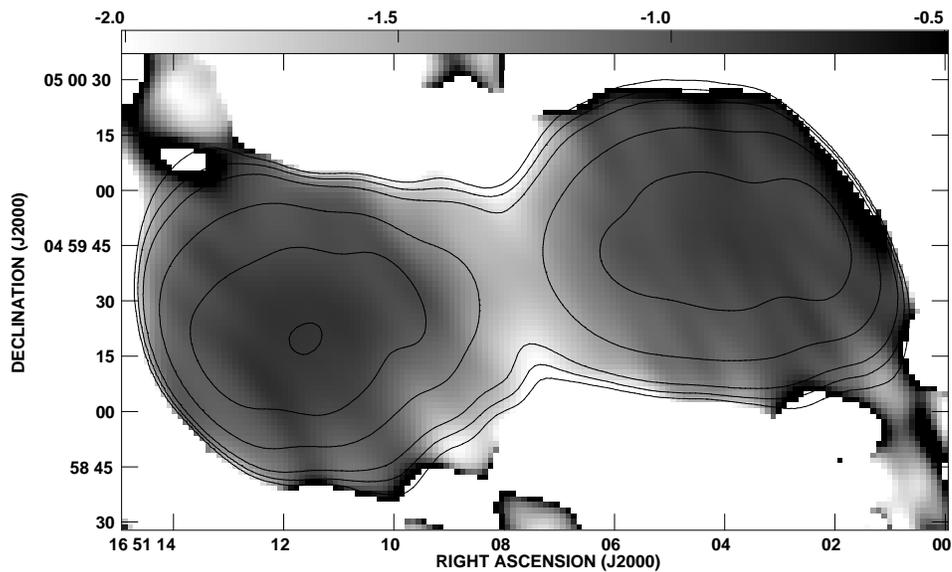}}

\end{picture}
\caption[]{Grey-scale of the spectral index map between 74 and 325~
MHz at 25.12 $\times$ 9.75~arcsec resolution with PA = $30.95^\circ$.
The grey-scale runs from $-2.5 \le \alpha \le -0.5$. Contours are of
the 325~MHz map at the 74~MHz resolution, separated by factors of 2
starting at $\simeq$ 81 mJy beam$^{-1}$. }

\label{74to325spix}
\end{figure*}

We present here the resulting spectral index map from our observations
at low frequencies, leaving however the more sophisticated, detailed
analysis for a subsequent paper by Gizani et al. in preparation as
stated above. Fig.~\ref{74to325spix} shows the spectral index map
between 74~ and 325~MHz, i.e. the image of $\alpha^{325}_{74}$. We
adopt the convention for spectral index $\alpha$, flux density
$S_{\nu}\propto \nu^{\alpha}$, where $\alpha <$ 0.

From this map, the average spectral index of the source is $\alpha
\approx -1.1$. The image hints at the distinction between the bright
and diffuse structures noted by \citet{GL2003}.  The former have
flatter, younger, spectra and the latter, as well as the overall
'cocoon', have older, steeper spectrum. However the spectral index of
the jets and rings is contaminated by the lobe material superposed
along the line of sight, and so the apparent spectral index of the
high-brightness features is steeper than the true value.

\citet{GL2003} interpreted the spectral index differences at high
frequency in terms of multiple outbursts which assume spectral
curvature.  This outburst interpretation is confirmed by our low
frequency observations, because the spectral differences between
different regions of the radio emission are smaller at the
$\alpha(74,325)$ map. The spectrum of the inner region curves, as
otherwise there would be a lot of emission with $\alpha  < -2.0$
(c.f. \citet{GL2003}).

The steep spectrum core found by \citet{GL2003} is hinted in the 325
 MHz map (see Fig.~\ref{Q}) and it is basically absent in
Fig.~\ref{74pt} at 74~MHz. This absence is not really a resolution
problem, since convolving our high resolution stacked map at 1.4~GHz
(see \citealt{GL2003}) with the 325~MHz beam, still shows the
presence of the core. Since the compact core is self-absorbed, we
suggest that its subsequent dissapearance at lower frequencies is
because of spectral curvature occuring at these frequencies. 

The relation between the turnover frequency $\nu_{break}$ (in GHz) 
versus the projected 
linear size $\theta$ (in mas) in a homogeneous, self-absorbed, incoherent 
synchrotron radio source, when the turnover in the spectrum is due 
to synchrotron self-absorption, is given by \citet{dea97}:

%\begin{equation}

\[
\theta \simeq 13.45\, (\frac{S^{2} B (1+z)}{\nu_{break}^{5}})^\frac{1}{4}
\] 

%\end{equation}

where $B$ is the magnetic field in Gauss, $S$ is the flux density at the 
peak in Jy, and $z$ is the redshift. For the Hercules A cluster the central 
magnetic field is of the order of $\mu$G 
($3 \lta B_{\circ}(\mu G) \lta 9$, \citet{GL99} 
and in preparation). Assuming that the turnover in the spectrum occurs at
 74 MHz where the flux of the compact core is $\simeq$ 1090 mJy, 
then its size $\theta$ is estimated to be $\geq$ 10 mas.  
Gizani,  Garrett \& Leahy, 2002 and in preparation, using new EVN and MERLIN
 observations of the core region at 18~cm, revealed emission coming 
from $ 10\times$ 20 mas scales (20 $\times$ 50 pc). 

\subsection{The Radio/X-ray interaction}

The analysis of the ROSAT PSPC data suggested that the X-ray emission
extends well beyond the radio emission \citep{GL2004}.  
%Fig.~\ref{hri}
%shows the HRI map smoothed with a gaussian of FWHM 10 arcsec, giving
%an effective resolution of 11.5 arcsec including the instrumental
%PSF. HRI detects with higher sensitivity X-ray emission from the inner
%region of the cluster, while the PSPC detector defines accurately the
%outskirts of the X-ray emission. 
The total mass of the cluster at 500
arcsec radius is $1.5\times 10^{14} M_{\odot}$, and the accretion rate
is $\dot{M} \simeq 15 M_{\odot}$ yr$^{-1}$ (adopting the relation of
the cooled mass which should be equal to the heat loss rate over the
heat per unit mass \citealt{IMetal}).

We expect the low frequency radio emission to originate, most likely,
from the oldest electron populations.  Our 74~MHz map could clarify if
there is any association between the X-ray and radio emission shown as
holes (areas empty from X-ray gas) or bright patches (cooler X-ray
emitting clouds of gas swept by the radio lobes), or other features
pointing to the fossil radio lobes, currently devoid of energetic
electrons. However there are no extra regions of radio emission at
74~MHz, so there seems to be no such interaction. Gizani \& Leahy,
2004 have shown that the low signal-to-noise ratio HRI data
and the low resolution PSPC data are insufficient to reveal a sign of
 interaction. Our high and low frequency radio observations on
Hercules A together with the analysis of the higher resolution and 
deeper observations by {\em Chandra}  
should help clarifying this matter (Wilson et al., 2005, in prep.).

%\begin{figure*}

%\centering
%\setlength{\unitlength}{1cm}
  
%\begin{picture}(17.5,12.0) 
%\put(-2,13){\special{psfile=./XRAY.PS
%		   hoffset=0 voffset=0 hscale=70 vscale=70 angle=270}}
%\end{picture}

%\caption{The HRI map of the Her~A cluster. The image has been smoothed
%with a gaussian of Full Width at Half Maximum (FWHM) $FWHM = 10$
%arcsec, giving an effective resolution of 11.5 arcsec including the
%instrumental PSF.  Contours are separated by factors of $\sqrt{2}$
%starting at $2.37 \times 10^{-9}$ W\,m$^{-2}$\,sr$^{-1}$ $= 3.07$ ct
%beam$^{-1}$ ($3\sigma$ above the background). The background level has
%been subtracted.}

%\label{hri}
%\end{figure*}

\section{Conclusions}
\label{concl}

We reported on the high resolution VLA observations of Her~A employing
the newly established VLA-PT link at 74~MHz.  We have also presented
the new A+B configuration map of the source at 325~MHz. The Her~A
image at 74~MHz with PT is dramatically different than the one
excluding the link. The bright ring-like and helical features in the
western and eastern lobe are resolved at the high resolution of
$\simeq 10 $~arcsec currently provided by the PT link.  These
structures were not even hinted in the earlier published map at the
same frequency.  In addition, the new images support the presence of a
complicated extended structure and a diffuse well-defined emission.
However the ends of the eastern and western lobes suffer from the
presence of ripples, probably due to residual self-calibration
problems, because the longer PT baselines do not overlap much with the
shorter ones. Another reason for the presence of these artifacts could
be the limitation in deconvolution induced by the fact that the $uv$
coverage for the PT link is relatively poor near the equator.

The spectral differences between different regions at high frequency
are smaller at low frequency. This is evidence of spectral curvature
confirming the multiple outburst interpretation used by Gizani and
Leahy (2003), to interprete the presence of regions with different age
(with younger and older material) apparent at the spectral index at
high frequencies.

The steep spectrum, self absorbed core of Hercules A is absent at
74~MHz. At 325~ MHz is barely visible. We suggest that its subsequent
dissapearance at lower frequencies is because of spectral curvature
occuring in these frequencies. We suggest that the likely candidate of 
the origin of the low frequency spectral curvature of the core is 
synchrotron self-absorption. 

Our low frequency data do not find any extra areas of radio emission
so the situation between the radio--X-ray correlation remains as
reported in Gizani \& Leahy (2004). 

Our long track observations for this exceptionally bright source will
provide critical input to the ongoing efforts to improve the
sophistication of VLA calibration for all low frequency observations,
and to extend their application to the PT link.  Our
observations will enlarge the currently short list of long,
uninterupted track data on bright objects at high resolution in order
to explore refractive, differential refractive, and related
ionospheric effects, and to gauge the algorithms' ability to correct
them on short time scales. Our VLA-PT link Her~A image can also be a
useful model for future calibration. 

\section*{Acknowledgments}

We acknowledge Paddy Leahy and Rick Perley for helpful discussions.
Nectaria Gizani would like to acknowledge the Naval Research
Laboratory travel and maintenance grant, with which her travel to the
US and data reduction was made possible. NG acknowledges the grant
PRAXIS XXI/BPD/18860/98 from the Funda\c c\~ao para a Ci\^encia e a
Tecnologia, Portugal for her post-doctoral fellowship of 2002--2003,
during which the observations and most of the data reduction of the
current work were made; Also the 2004--2007 post-doctoral grant SFRH /BPD/11551/2002,  
from the same Institute during which this work was published. 
NG also acknowledges the State Scholarships Foundation (IKY),
Greece, for her 2003-2004 post-doctoral grant under contract 332 during
which the current paper was written. Finally NG is grateful to the Departamento de F\'{i}sica 
of the Faculdade de Ci\^{e}ncias of the Universidade de Lisboa, Portugal for 
allowing her to use its installations to carry out this task. 
 
Basic research in astronomy at the NRL is funded by the Office of
Naval Research. The National Radio Astronomy Observatory is a facility
of the National Science Foundation operated under cooperative
agreement by Associated Universities, Inc.  This research has made use
of the NASA/IPAC Extragalactic Database (NED) which is operated by the
Jet Propulsion Laboratory, Caltech, under contract with the National
Aeronautics and Apace Administration.  This research has also made use
of NASA's Astrophysics Data System.

\bibliography{3c348}

\end{document}